\documentclass[aps,prl,twocolumn,superscriptaddress,showpacs, floatfix]{revtex4-1}
\usepackage{xcolor}
\usepackage[utf8]{inputenc}
\usepackage{amsmath}
\usepackage{amssymb}
\usepackage{graphicx}
\usepackage{graphics}
\usepackage[T1]{fontenc}

\begin{document}
\title{Theory of ultrafast magnetization of non-magnetic semiconductors with localized conduction bands}

\author{Giovanni Marini}
\email{giovanni.marini@iit.it}\affiliation{Graphene Labs, Fondazione Istituto Italiano di Tecnologia, Via Morego, I-16163 Genova, Italy}
\author{Matteo Calandra} 
\email{m.calandrabuonaura@unitn.it}
\affiliation{Department of Physics, University of Trento, Via Sommarive 14, 38123 Povo, Italy}
\affiliation{Sorbonne Universit\'e, CNRS, Institut des Nanosciences de Paris, UMR7588, F-75252 Paris, France}
\affiliation{Graphene Labs, Fondazione Istituto Italiano di Tecnologia, Via Morego, I-16163 Genova, Italy}
\email{m.calandrabuonaura@unitn.it}

\begin{abstract}
The magnetization of a non-magnetic semiconductor by femtosecond light pulses is crucial 
to achieve an all-optical control of the spin dynamics in materials and to develop faster memory devices. However, the conditions for its detection are largely unknown. 
In this work we identify the criteria for the observation of ultrafast magnetization and critically discuss the  difficulties hindering its experimental detection.
We show that ultrafast magnetization of a non magnetic semiconductor can be observed in compounds with very localized conduction band states and more delocalized valence bands, such as in the case of a p-d charge transfer gap. By using constrained and time dependent density functional theory simulations, we demonstrate  that a transient ferrimagnetic state can be induced in diamagnetic semiconductor V$_2$O$_5$ via ultrafast pulses at realistic fluences. The ferrimagnetic state has opposite magnetic moments on vanadium (conduction) and oxygen (valence) states.
Our methodology outruns the case of V$_2$O$_5$ as it identifies the key requirements for a computational screening of ultrafast magnetism in non-magnetic semiconductors.
\end{abstract}
\maketitle

The ultrafast all-optical control of the spin dynamics in materials via femtosecond light pulses is an extremely appealing perspective as it could lead to a new generation of memory devices much faster than the widespread magneto-optical storage media or random access memories\cite{Kimel2019,PhysRevLett.99.047601,10.1126/science.1253493}.
The investigation of magnetic phenomena at the femtosecond scale~\cite{doi:10.1021/jacs.9b10533} includes ultrafast demagnetization
\cite{PhysRevLett.76.4250}, namely the melting of the magnetic order of a material  by femtosecond pulses, magnetization reversal\cite{PhysRevLett.76.4250,PhysRevLett.99.047601}, i.e. the use of ultrafast laser pulses to invert the direction of the magnetization and the possibility to induce a phase transition from an antiferromagnetic phase to a ferromagnetic one\cite{PhysRevLett.93.197403}. Furthermore, the  enhancement of a pre-existent ferromagnetic order using ultrafast irradiation has been demonstrated in GaMnAs\cite{PhysRevLett.98.217401} and more recently in the layered compound Fe$_3$GeTe$_2$. In Fe$_3$GeTe$_2$ the Curie temperature has been enhanced and tuned by ultrafast laser pulses by manipulating the magnetic anisotropy energy and, consequently, the itinerant magnetic order of the material\cite{PhysRevLett.125.267205}. 

In all these cases, the material is assumed to be magnetic in its ground state and the magnetic state is altered via excitation. A different (yet related) phenomenon, of which there is still little evidence, is the ultrafast magnetization of a non-magnetic material via femtosecond light pulses. Ultrafast magnetism has been induced in nonmagnetic materials using circularly polarized light through inverse Faraday effect\cite{Cheng2020,PhysRevB.86.100405}, although the resulting magnetism is short lived and the non-magnetic nature of the irradiated material is not altered. The aforementioned experimental results demonstrate that in some conditions light can modify the magnetic properties of the material. It is then natural to wonder whether the ultrafast stabilization of a ordered magnetic state in a non-magnetic material via femtosecond pulses is a realistic possibility. Namely, by promoting electrons in the conduction band of a non-magnetic material and the consequent stabilization of a thermalized electron-hole plasma, is it possible to stabilize a magnetic transient state at the ultrashort timescale before electron-hole recombination takes place?
Finally,  under what conditions can this happen and what are the criteria to identify materials possessing this remarkable property? 

While at present there are still many open questions regarding the possibility to induce a magnetic state in a non-magnetic material via ultrafast light, it is clear that the ability to do so would open new perspectives in tailoring magnetic properties of materials. The goal of this work is to address these fundamental questions from a theoretical point of view. We give general criteria for the occurrence of ultrafast magnetization in non-magnetic materials and we target the layered diamagnetic semiconductor V$_2$O$_5$  as a system that can be 
made magnetic via ultrafast light pulses.


For the sake of simplicity let us consider a non-magnetic semiconducting or insulating compound with a gap of the order of the eV or larger. This assumption is needed to ensure that the electron-hole recombination rate is long enough, namely of the order of nanoseconds. By using ultrafast laser pulses a substantial number of electrons \cite{footnotecarriers} can be promoted in the optically active conduction band and an electron-hole plasma  (hole in the valence band and electrons in the conduction band) is induced in the system. At large enough  photocarrier concentrations the excitons are screened and the electron-hole plasma thermalizes fairly quickly, below the picosecond scale, so that the system effectively feels an unbalanced population of carriers that can be described by two Fermi distributions, one for the holes and one for the electrons\cite{PhysRevLett.82.4340,PhysRevB.65.054302,PhysRevB.104.144103}.
The question we want to answer is under what circumstances this transient state occurring before recombination is magnetic.

Ultrafast magnetization can happen via several material specific mechanisms (e.g. enhancement of the magnetic anisotropy energy in a 2D material), however a general one is the following. If the conduction band states have very different localization properties than the valence band states, as it happens in the case of a p-d gap (p-states in valence and d-states in the conduction bands) or in the case of a weakly dispersive conduction band (flat band) and a strongly dispersive valence band, then the electron-electron interaction felt by the excited electrons is very different from the one in the ground state. As a consequence a magnetic instability can in principle occur in the transient state. 

\begin{figure*}[t!]
\centering
\includegraphics[width=1\linewidth]{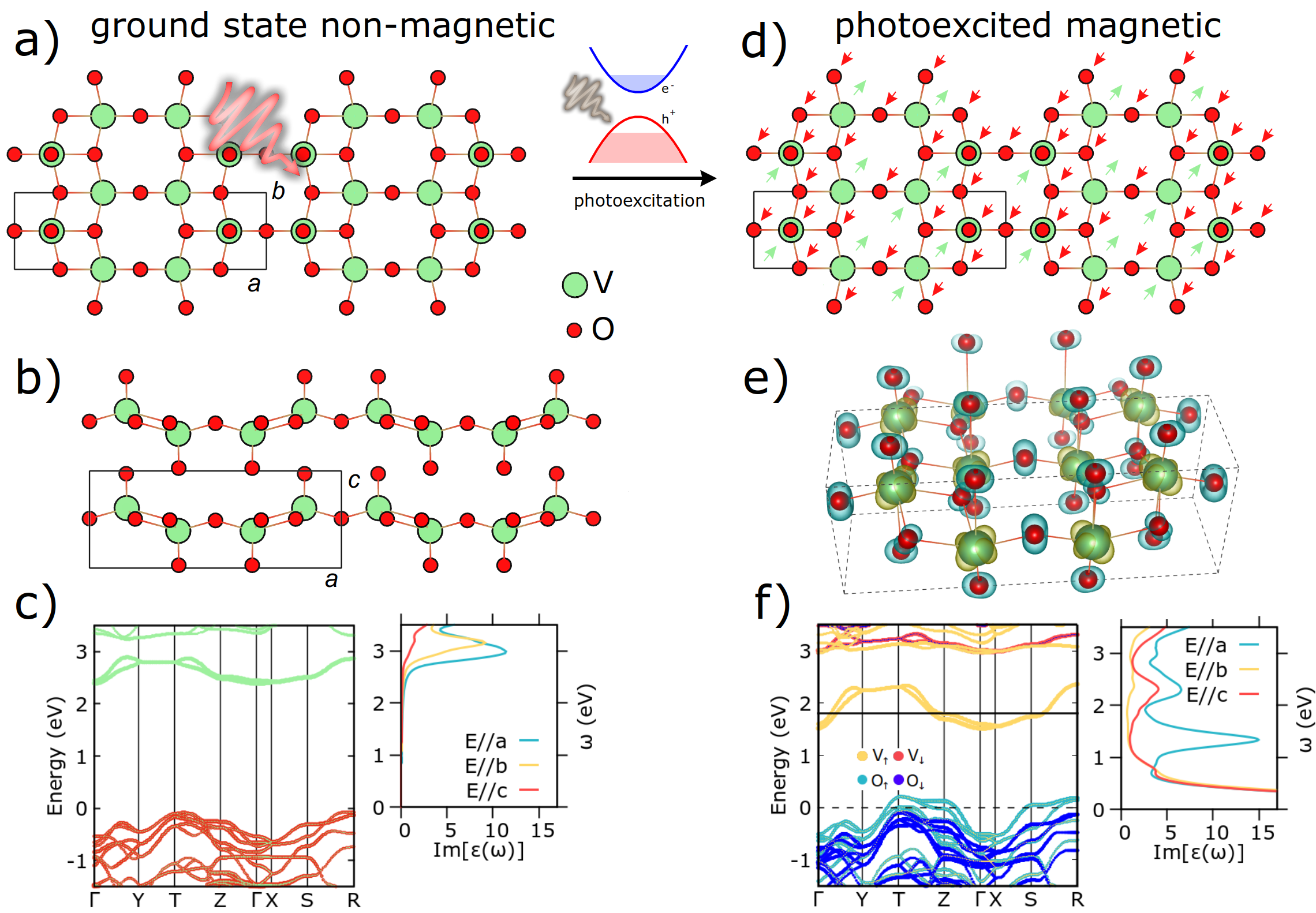}
\caption{Light induced magnetism in V$_2$O$_5$. Panel~a,b): V$_2$O$_5$ in the $\alpha$ structure before irradiation (top and lateral views). Panel~c): ground state electronic structure of V$_2$O$_5$ including the projection onto atomic vanadium (green) and oxygen (red) atomic orbitals. Panel~d): ferrimagnetic order stabilized by ultrafast radiation. Panel~e): three-dimensional plot of spin-polarized charge density in the photoexcited state. Yellow (cyan) regions represent spin up (down) charge excess.   Panel~f): Electronic structure of $\alpha$-V$_2$O$_5$ for $n_e = 0.25~e^-/f.u.$, projected onto spin-resolved vanadium (green) and oxygen (red) orbitals. The imaginary part of the dielectric function resolved in the three spatial direction for the ground state (panel~a) and in the presence of an electron-hole plasma (panel~d) is plotted on the right hand side of the corresponding electronic structures (panels~c and f).}\label{fig1}
\end{figure*}

We give a practical demonstration of our idea for the van-der-Waals compound vanadium pentoxide (V$_2$O$_{5}$). This compound is mainly studied for its performances as cathode material in the realization of batteries\cite{YAO2018205}. V$_2$O$_{5}$ has a rich structural phase diagram with several competing phases\cite{doi:10.1021/acs.inorgchem.8b01212}. The most commonly stabilized structure is the $\alpha$-structure (space group 59, $Pmmn$) and is characterized by one-dimensional ladders along the y-direction, depicted in Fig.~\ref{fig1}. Electronically, the material behaves as an insulator, due to the presence of an indirect band gap between the valence and the conduction band. The optical band-gap is $\approx$ 2.35 eV\cite{Hieu1981}, however different values in the 2.2-2.8 eV range have also been reported\cite{PhysRevB.41.4993,Chain:91,doi:10.1063/1.3611392}. Experimentally V$_2$O$_5$ is non magnetic, however, antiferromagnetism can be induced by Na doping\cite{CARPY1972229} and 
NaV$_2$O$_5$ has a N\'eel temperature of $320\pm50$ K, in agreement with  theory\cite{PhysRevB.92.125133}. 
Due to its layered nature, it has been possible to synthesize few layer V$_2$O$_5$ via liquid exfoliation, as well as supported V$_2$O$_5$/TiO$_2$ monolayers\cite{MACHEJ1991115}. Some experimental ultrafast investigations on V$_2$O$_5$ are already present in literature, including a study of carrier relaxation in V$_2$O$_5$ nanowires\cite{doi:10.1063/1.4823506} and the possibility to synthesize VO$_2$ from V$_2$O$_5$ via femtosecond pulsed laser deposition\cite{doi:10.1063/5.0010157}.  

We simulate the ground state and excited states properties within density functional theory (DFT) by using the Quantum ESPRESSO (QE) package\cite{QE,QE2}. The excited electron-hole plasma is modeled by
using constrained density functional perturbation theory (cDFPT) that we recently developed \cite{PhysRevB.104.144103} extending previous theoretical results\cite{PhysRevB.65.054302}. This amounts to suitably constrain the occupations of the Kohn-Sham conduction eigenstates in order to mimic the thermalized photocarrier population. In more details, electrons are removed from the oxygen valence states (leaving behind a thermal distribution of holes) and are added to the conduction bands. All structural and electronic properties are then calculated by constraining two thermal distributions of electrons and holes in the conduction and valence band, respectively.  
We include  correlation effects  employing the $GGA+U$ approach\cite{PhysRevB.71.035105} (including the Hubbard parameters on the vanadium d states). In order to maintain a completely $ab$-$initio$ approach, the $U$ value is determined self consistently from first principles, using the method introduced by Timrov $et$ $al.$\cite{PhysRevB.98.085127,PhysRevB.71.035105} that we modified to include the presence of an electron-hole plasma. Thus, we calculate the on-site repulsion $U$ parameter both in presence and absence of the electron-hole plasma, finding in both cases  very similar values of $U\approx4.4$ eV. Further technical details are given in the Supplemental Material\cite{SM}, which includes Refs.\cite{doi:10.1021/acs.jctc.6b00114,QE,QE2,PBE,PhysRevB.71.035105,MP,PhysRevB.98.085127,PhysRevB.104.144103,https://doi.org/10.1002/jcc.20495,GROSSO2000425,Lamsal2013,PhysRevB.92.125133,doi:10.1021/acs.inorgchem.8b01212,Elliott2018,PhysRevLett.52.997,doi:10.1063/1.5142502,PhysRevB.58.3641,doi:10.1021/ct800518j,PhysRevB.98.094306,doi:10.1063/1.4921690,Miyamoto2021,doi:10.1021/acs.jctc.5b00621}.

In Fig.~\ref{fig1}, panels~a) and b) we report the structure of $\alpha$-V$_2$O$_5$. Each vanadium atom is surrounded by five oxygen atoms: one oxygen is exclusively bound to the vanadium, three of them are bound to three vanadium atoms in the same ladder, while one oxygen behaves as a spacer between ladders and is shared between two vanadium atoms belonging to different ladders. The ground state electronic structure is shown in Fig.~\ref{fig1} c). It displays an indirect band gap of 2.47 eV, in good qualitative agreement with the experimental value of 2.35 eV\cite{Hieu1981}. The projection of the Kohn-Sham states onto atomic vanadium and oxygen orbitals is also shown and the size of the dots is proportional to the atomic character. Projections over atomic orbitals demonstrate that valence bands are mainly composed by oxygen($p$)-derived states, while vanadium($d$) states mainly constitute the conduction band. Furthermore, the narrow bandwidth of the conduction band should also favour the formation of a correlated state.

We then perform constrained DFT calculations for several photocarrier concentrations (PCs) in order to investigate how the presence of an electron-hole plasma affects the electronic structure of this compound. We find that ultrafast magnetization  can occur in this material, as shown in Fig.~\ref{fig1} where panel~d) schematically represents the magnetic ordering observed after irradiation.
Magnetism is characterized by the vanadium and oxygen sub-lattices presenting opposite magnetic moments. The magnetic state can be well understood by looking at the spin up minus spin down charge plot in Fig.~\ref{fig1} e). Here, yellow surfaces represent a spin up excess charge while the blue surfaces represent spin down excess charges. The alternating magnetic moments of the vanadium and oxygen sub-lattices
leads to a transient ferrimagnetic order after irradiation, resulting in a very weak (close to zero) total magnetic moment. 

The effect of the electron-hole plasma on the electronic structure is shown in Fig.~\ref{fig1} f), where the Kohn-Sham eigenvalues for $\alpha$-V$_2$O$_5$ are plotted for a PC value $n_e = 0.25~e^-/f.u.$. The excitations are of the charge transfer type from the O-valence band to the V-conduction band. The ferrimagnetic state is signalled by  the non-degeneracy of the up and down bands. The observed ferrimagnetic configuration can be understood thinking to the simultaneous effect of hole and electron doping. This is confirmed by performing calculations in a rigid doping approach (see Supplemental Material) demonstrating that hole and electron doping produce different magnetization and magnetic moments residing on vanadium (oxygen) atoms for electron (hole) doping.

 In panels~c) and f), we also show the imaginary part of the dielectric function corresponding to the the ground state and photoexcited electronic structures, respectively. The dielectric function is resolved into the three spatial directions. For the ground state, we find that our calculations are in qualitative agreement with optical measurements\cite{Lamsal2013} (see Supplemental Material), thus justifying our procedure. Laser irradiation completely changes the optical properties of the system. In the low-energy region (0-1.5 eV)  a metallic-like contribution appears due to intraband scattering and a reduction of the bandgap caused by irradiation occurs. As a consequence, the prominent peak observed at $\approx$ 3 eV in panel~c) for the incident field parallel to the $a$ direction (cyan curve) is shifted towards lower energies in panel~f)( $\approx 2.3$ eV). Since the reduction of the gap only happens if the system becomes magnetic after irradiation (see Supplemental Material where we simulate the optical properties  of photoexcited V$_2$O$_5$ neglecting magnetic ordering), optical measurements can be seen as a fingerprint to directly probe the irradiation-induced magnetization in V$_2$O$_5$. 

\begin{figure}[t!]
\centering
\includegraphics[width=0.85\linewidth]{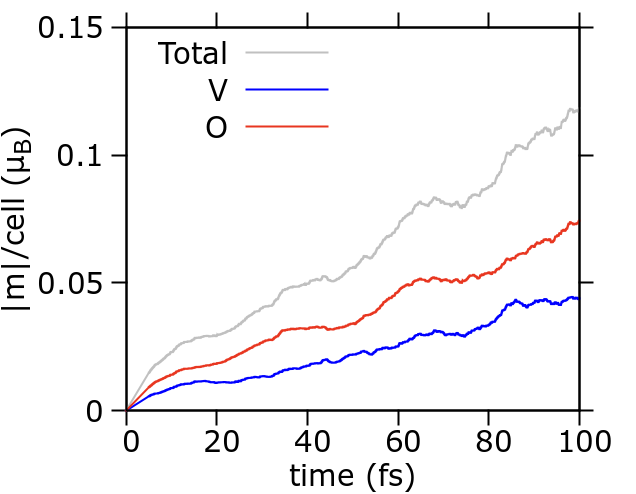}
\caption{ Light induced magnetization dynamics in TDDFT: grey curve represents the absolute magnetic moment per unit cell (a unit cell includes two formula units), red and blue curves represent the separated contribution of oxygen and vanadium atoms to the absolute magnetic moment, respectively.}\label{fig2}
\end{figure}

The discussion presented here pertains to a quasi-equilibrium situation where both electrons and holes have reached an intraband thermalization via carrier-carrier scattering. Nevertheless, constrained DFT simulations do not allow to describe the evolution towards a magnetic state occurring in the first femtoseconds after irradiation. In order to answer to this question, we performed  time-dependent density functional theory (TDDFT) simulations. In Fig.~\ref{fig2}, we plot the absolute magnetic moment as a function of time for photoexcited V$_2$O$_5$, as obtained in TDDFT: an ever increasing absolute magnetic moment is observed in the first 100 fs, reaching a value greater than 0.1 $\mu_B$/unit cell. The TDDFT results, extensively discussed in the Supplemental Material, demonstrate that a finite absolute magnetic moment develops in the compound following the laser irradiation and that a fundamental role is played by the spin-orbit interaction. We conclude that no dynamical barrier between the ground state and the quasi-equilibrium magnetic state predicted using cDFPT is present, and thus that the photoexcited magnetic state can be experimentally realized.

\begin{figure*}[t!]
\centering
\includegraphics[width=1\linewidth]{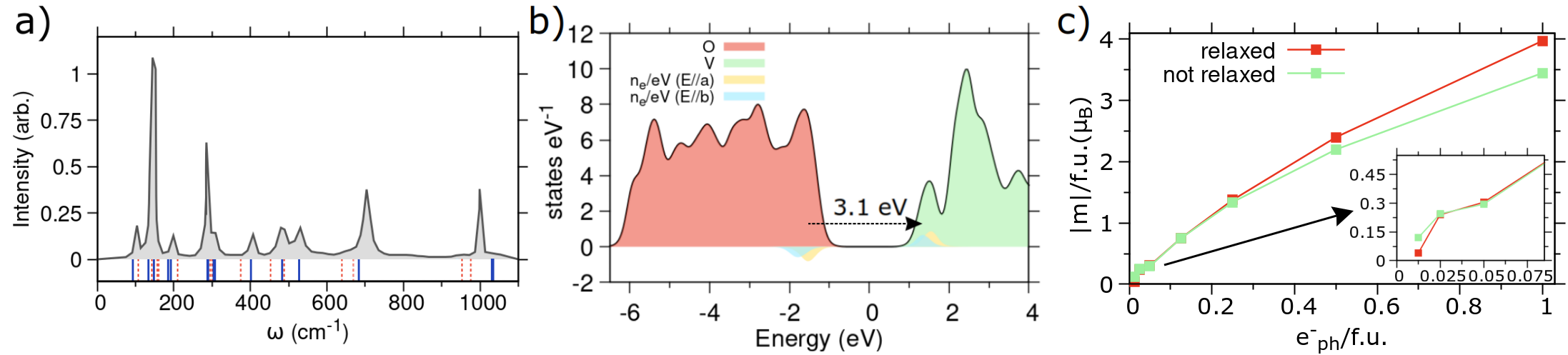}
\caption{Characterization of photoexcited  V$_2$O$_5$. Panel~a): Calculated Raman frequencies for $\alpha$-V$_2$O$_5$ (blue lines) compared with the experimental Raman spectrum\cite{https://doi.org/10.1002/jrs.5616} (grey shaded region) and calculated photoexcited Raman frequencies at $n_e = 0.25~e^-/f.u.$ (red dashed lines). Panel~b): density of states (red and green indicate filling indicate valence and conduction, respectively) and corresponding energy-resolved photocarrier concentration (light blue and yellow filling) induced by a 3.1 eV laser having a fluence of 5 mJ/cm$^2$. Panel~c): absolute magnetic moment vs photocarrier concentration with (red) and without (green) performing structural optimization with respect to the unit cell parameters.}\label{fig3}
\end{figure*}

The Raman frequencies of the ground state structure and at a photocarrier concentration of $n_e = 0.25~e^-/f.u.$
are shown in Fig.~\ref{fig3} a) (blue lines for the ground state). For the ground state, we find a good agreement with previously published theoretical and experimental results\cite{https://doi.org/10.1002/jrs.5616,doi:10.1021/acs.inorgchem.8b01212}. The grey filled curve are experimental data from Ref.\cite{https://doi.org/10.1002/jrs.5616}. Dashed red lines label Raman frequencies after photoexcitation. 
We observe a general softening of the high frequency modes (above 400 cm$^{-1}$), together with a general hardening of the modes in the range 0-400 cm$^{-1}$. 
This prediction and the optical features discussed before are the fingerprints
of the photoexcited ferrimagnetic state. 

While the physical model presented here is conceptually simple, some caveats are needed. First of all, the fluence of the ultrafast pulse must be low enough so that the energy transferred to the sample in the region illuminated by the laser  is smaller than $k_B T_c$, where $k_B$ is the Boltzmann constant and $T_c$ is the Curie  temperature (i.e. the magnetic temperature below which magnetism can occur). Moreover, the different localization properties of the valence and conduction bands often imply that the excited electrons feel very different forces from the lattice at scale larger than the picoseconds. As a consequence, a structural phase transitions could in principle occur with unpredictable consequences on the magnetic properties\cite{PhysRevLett.122.145702}. If, however, these two  difficulties are avoided, then transient ultrafast magnetization can occur. 

We discuss these two points for the case of V$_2$O$_5$. The occurrence of a ferrimagnetic state could be hindered by a low Curie temperature so that the energy transferred to the sample by the laser overrules $k_B T_c$ even at low fluences. Despite the fact that it is difficult to obtain a reliable estimation of the Curie temperature from first principles, especially in the case of a complex system like $\alpha$-V$_2$O$_5$, we can try to compare our situation with the one of electron doped V$_2$O$_5$. In V$_2$O$_5$ electron doping can be achieved by creating oxygen vacancies\cite{doi:10.1063/1.4899249} that leaves uncompensated electrons in the vanadium d$_{xy}$ states, an effect well described by first principles calculations\cite{doi:10.1063/1.3146790}. The presence of oxygen vacancies would cause the appearance of a ferromagnetic state in V$_2$O$_{5-x}$ for x $<$ 0.13 or 0.19 $<$ x $<$ 0.45\cite{doi:10.1063/1.3146790}. Experimentally, the total magnetic moment is found to be unrelated to the type of vacancies (there are several non equivalent oxygen sites) and the ferromagnetic Curie temperature is fairly above room temperature\cite{doi:10.1063/1.4899249}. We infer that similar values for T$_c$ can be expected in our system , as the photoexcited magnetic state reported here and the magnetic state originating from electron doping possess a common origin and a total magnetic moment of the same order is calculated for both cases (see Supplemental Material).

We now evaluate the energy transferred to the crystal by the laser light as a function of the photocarrier concentration (i.e. as a function of the fluence of the incoming laser) to understand if magnetism is possible.
In Fig.~\ref{fig3} b) we calculate the maximum plasma electron-hole density generated by a 3.1 eV laser having a fluence of 5 mJ$/$cm$^2$ with in-plane polarized light (E $\perp$ c) (see the Supplemental Material for more details). Averaging between the two planar polarization directions, we calculate that an electron-hole density as high as $0.22~e^-/f.u.$ can be generated in the first layer of the sample. The plot in Fig.~\ref{fig3}~c) demonstrates that magnetization sets in much earlier, namely at $0.015~e^-/f.u.$, corresponding to $1.35~\times~10^{20}~e^-/$cm$^3$. Here, the red and green curves represent the absolute magnetic moment calculated with and without performing structural optimization with respect to the unit cell parameters in the cDFT calculation, respectively. The absolute magnetic moment grows linearly for $n_e \leq 0.25~e^-/f.u.$. For higher electron densities, we observe a sub-linear increase of the absolute magnetic moment in both cases. We find that a photocarrier concentration of $0.22~e^-/f.u.$ corresponds already to an absolute magnetic moment in excess of 1~$\mu_B/f.u.$~. The comparison between the two curves shows that structural changes contribute to enhance the photoinduced magnetization.

The 3.1 eV energy absorbed by each photocarrier corresponds to an excess energy of 0.63 eV with respect to the gap. This translates to an excess energy $\Delta E = 0.14 $~eV$/f.u$ at  $n_e = 0.22~e^-/f.u.$, which is mostly transferred to the lattice via intraband thermalization (here we neglect the energy transfer to the spin subsystem). Combining these findings with experimental heat capacity data\cite{doi:10.1021/ja01295a006}, we now estimate the $quasi$-equilibrium temperature for the system (before carrier recombination takes place). Assuming an initial temperature $T=50$~K and neglecting energy dissipation, we estimate a $quasi$-equilibrium temperature T$_f=168.5$~K at $0.22~e^-/f.u.$~ ( T$_f=250$~K if we consider $n_e = 0.5~e^-/f.u.$~). Both these temperatures are lower than room temperature, supporting the possibility to experimentally observe ultrafast magnetization in V$_2$O$_5$.

We now discuss the structural stability of the photoexcited phase. As we stressed before, injection of a large number of photocarriers can trigger reversible or irreversible phase transitions due to the possible population of antibonding states resulting in strong forces on the ions. For this reason it is important to calculate the structural and vibrational properties of the photoexcited transient phase. We find that the transient photoexcited ferrimagnetic state is dynamically stable, $i.e.$ no imaginary phonon frequencies are observed (see Supplemental Material). Interestingly, another crystal structure structure, commonly referred to as $\beta$-V$_2$O$_5$ ($P2_1m$, space group 11), becomes the lowest energy structure under photoexcitation. However, as imaginary phonons are absent from $Pmmn$ phonon spectrum, an energy barrier exists between the two structures. We calculate a very large energy barrier of $1.25$ eV/atom by interpolating between atomic coordinates of the two structures. Thus, we do not expect any structural transition to occur under photoexcitation. 

In conclusion, we identify the general criteria for ultrafast magnetization of non-magnetic semiconductors to occur. The key requirement is that the conduction band should host very localized states or flat bands while the valence more delocalized ones, like in the case of a p-d charge transfer gap. Moreover, we discuss the possible difficulties hindering the experimental detection of this effect. By using constrained density functional perturbation theory calculations, we demonstrate that femtosecond light pulses can induce a magnetic state in an otherwise non-magnetic material, namely the diamagnetic semiconductor V$_2$O$_5$. Our calculations show that a transition from the ground state non-magnetic state to a ferrimagnetic state develops upon laser irradiation, even for very low plasma densities. Thus, light can shape magnetic properties, notably not just suppressing an existing magnetic phase, but also inducing a new magnetic phase. 
Our methodology outruns the case of V$_2$O$_5$ and paves the way to a computational screening of semiconducting materials displaying ultrafast magnetization.

Finally, we point out that, while our conditions for the occurrence of ultrafast magnetization are very general,  in specific materials, such as 2D materials, other effects could cooperate in inducing the magnetic order. For example, as the Mermin and Wagner theorem forbids the breaking of a continous symmetry in 2D, any mechanism enhancing the magnetic anisotropic energy and making the photoexcited magnetic system more Ising like could lead to an enhancement of magnetic order in a 2D or a quasi 2D material. This is the mechanism claimed to be at the origin of the enhancement of magnetic order in Fe$_3$ GeTe$_2$ by ultrafast light pulses.

\begin{acknowledgments}
The authors acknowledge support from the European Union's Horizon 2020 research and innovation program Graphene Flagship under grant agreement No 881603. We acknowledge the CINECA award under the ISCRA initiative and PRACE, for the availability of high performance computing resources and support.
\end{acknowledgments}

\bibliography{bibliography}
\bibliographystyle{apsrev4-2}
\end{document}